\documentclass[a4paper,12pt,epsf]{iopart}
\usepackage[dvips]{graphicx}
\usepackage{amssymb}
\usepackage{amsfonts}
\newcommand{\be}{\begin{equation}}
\newcommand{\ee}{\end{equation}}

\newcommand{\bea}{\begin{eqnarray}}
\newcommand{\eea}{\end{eqnarray}}
\newcommand{\beastar}{\begin{eqnarray*}}
\newcommand{\eeastar}{\end{eqnarray*}}

\newcommand{\ds}{\displaystyle}
\newcommand{\eq}[1]{~(\ref{#1})}

\newcommand{\order}{{{\cal O}}}

\newcommand{\eps}{\epsilon}
\newcommand{\teq}{t_{\rm eq}}

\newcommand{\rate}{\Gamma}

\newcommand{\xbar}{\bar{x}}
\newcommand{\p}{P}

\newcommand{\ddtau}{\frac{\partial}{\partial \tau}}
\newcommand{\height}{h}
\newcommand{\logtime}{\nu}
\newcommand{\ptilde}{\tilde{P}}

\usepackage{epsfig}
\newcommand{\dbar}{\bar{d}}
\begin{document}

\title{Anomalous coarsening and glassy dynamics}
\author{M. R. Evans\dag
\footnote[3]{m.r.evans@ed.ac.uk}}
\address{
\dag\ Department of Physics and Astronomy,\\ University of Edinburgh,
       Mayfield Road, Edinburgh EH9 3JZ, U.K. }

\begin{abstract}
  An overview of the related topics of anomalous coarsening and glassy
  dynamics is given.  In anomalous coarsening, the typical domain size
  of an ordered phase grows more slowly with time than the power law
  dependence that is usually observed, for example, in magnetic
  systems.  We discuss how  anomlaous coarsening may arise through
  domain-size dependent energy barriers in the coarsening process.  We
  also review the phenomenology of glassy dynamics and discuss how
  simple nonequilibrium models may be used to reproduce certain
  aspects of the phenomenology. In particular, models involving
  dynamical constraints that give rise to anomalous coarsening are
  considered. Two models, the Asymmetric Constrained Ising Chain and
  the ABC model, are discussed in detail with emphasis on how the
  large energy barriers to coarsening arise through the local
  dynamical constraints. Finally, the relevance of models exhibiting
  anomalous coarsening to glassy systems is discussed in a wider
  context.
\end{abstract}
\maketitle
\section{Introduction}
In this paper I will review some very simple dynamical models whose
dynamics slow down with time. Along with this, some typical length
scale in the system grows.  These two features are referred to as
`glassy dynamics' and `coarsening' respectively. Anomalous coarsening
refers to the situation where a length scale grows more slowly with
time than the usual power law dependence.  In this introduction I will
briefly review these topics and their interconnections.  Then in the
later sections I will describe in detail two models that exhibit
anomalous coarsening and glassy dynamics.  The appealing feature of
the models to be discussed is that although they exhibit non-trivial
behaviour they are simple enough to analyse and gain a firm
understanding of.  In Sections~2,3 I will summarise these analyses.
Although the choice of models to be reviewed reflects personal
research interests, I believe that the two models studied in
Sections~2 and 3 are each representative of a class of systems. To
bring this out I try to make connections to other related models in
Sections 2.5 and 3.8.  Finally, in Section~4, I return to the relation
between glassy dynamics and coarsening.

\subsection{Glassy dynamics}
The phenomenolgy of glassy systems is well established---see
\cite{Glasses} for excellent reviews. Experimentally, the archetypal
system is a liquid that, when cooled rapidly to temperatures where the
equilibrium state should be a crystalline solid, becomes trapped in
metastable liquid-like configurations. Thus the higher temperature
configurations are frozen in and one can meaningfully say that the
glass is like a frozen liquid\footnote{Strictly one may distinguish
  between a supercooled liquid and a glass according to the rate of
  the cooling schedule but this issue is not pertinent here.}.

Three distinguishing features of the glassy state are the long
relaxation times, stretched exponential decay of correlation functions
(see equation~3 below) and ageing phenomena\cite{BCKM} whereby since the
system is out of thermal equilibrium it evolves continuously as time
goes by and time-translational invariance is lacking.  This
phenomenology provides a mandate for theoretical study.

The long relaxation times, that show a non-Arrhenius divergence
as the temperature $T$ is lowered are often fitted experimentally
by the Vogel-Tammann-Fulcher (VTF) law
\begin{equation}
\tau = \tau_0\exp[-\mbox{B}/(T-T_0)]\;.
\label{VTF}
\end{equation}
The relaxation time $\tau$ may characterise, for example,
the time for an externally imposed stress to relax.
Although some heuristic justifications have been offered \cite{AG},
for practical purposes VTF is just a fit with three parameters
$\tau_0, B, T_0$.  $T_0=0$  reduces to an Arrhenius law.  A system
for which $T_0$ is small, so that one has something close to Arrhenius
behaviour, is referred to as a `strong glass', whereas a system
exhibiting large deviations from Arrhenius behaviour is referred to as
a `fragile glass'.  Generally, $T_0$ is much lower than the
experimental temperatures so although there is a singularity in the
fit, it is not physically relevant.  Nevertheless it should be noted
that there has been a long debate concerning whether $T_0$ represents
a true thermodynamic transition temperature achievable in the limit of
infinitely slow cooling.

On the other hand, other alternative functional forms for 
relaxation  times $\tau(T)$
have been proposed that do not exhibit singularities at any finite
$T$. Among these, the exponential inverse temperature squared (EITS)
form
\begin{equation}
\tau\sim \exp( \mbox{const}/T^2)
\end{equation}
(where the constant is positive) is popular.  Experimentally, it is
difficult to distinguish between VF and EITS behaviour due to obvious
limitations on the longest accessible timescales; both can represent the
experimentally observed $\tau(T)$ in many
materials~\cite{RB}.

Stretched exponential decay of correlation functions, let us say an
autocorrelation $q(t)$, is expressed by the Kohlrauch-Williams-Watt
law
\begin{equation}
q(t) \sim \exp\left[ - (t/\tau)^\theta\right]
\label{strexp}
\end{equation}
where the stretching exponent $\theta <1$.  An heuristic explanation
for this law is to postulate a broad distribution $\Omega(\tau)$ of
relaxation modes with decay constants $\tau$ yielding
\begin{equation}
q(t) = \int d\tau\ \Omega(\tau) \exp( - t/\tau).
\end{equation}
If one assumes $\Omega(\tau) \sim \exp(-a \tau^\gamma)$ then for large $t$ a poor
man's saddle point argument implies that the dominant modes have
$\displaystyle \tau= (t/a\gamma)^{1/(\gamma+1)}$ which leads to
(\ref{strexp}) with $\theta = \gamma/(1+\gamma)$.  However the
question remains as to how the broad distribution of modes
$\Omega(\tau)$ comes about.

\subsection{Kinetically constrained models}
\label{KinCon}
One idea that was proposed to generate a broad distribution of
relaxation times was of a hierarchy of degrees of freedom\cite{PSAA}.
The different levels in the hierarchy then relax in serial, the
degrees of freedom in one level having to wait for the degrees of
freedom in the level above to reach some configuration before they are
free to evolve. This latter condition is a realisation of a 
{\it dynamical constraint}.

A more concrete realisation of a dynamical constraint in a system with just one
set of degrees of freedom is the $n$-spin facilitated kinetic Ising
models introduced by Fredrickson and Anderson \cite{FA}-\cite{PYA}.
That model comprises non-interacting Ising spins in a downwards
pointing field. However a spin can only flip if at least $n$ nearest
neighbour spins are pointing up (against the field). This gives rise
to slow cooperative relaxation.  A modification of this model is to
have anisotropic dynamical constraints \cite{JE}--\cite{SE}.  In the
one-dimensional version which we will refer to as the Asymmetric
Constrained Ising Chain (ACIC) a spin can flip down only if its left
neighbour is pointing up.  As we shall show in Section~\ref{ACIC},
although
the equilibrium distribution is the Boltzmann distribution, the
relaxation to equilibrium is strongly affected by the asymmetric
constraint. Thus when quenching from high temperature to low
temperature the equilibrium distribution implies that most spins
should be pointing down. However for a spin to flip down its neighbour
has to point up. Thus there is an energy barrier for isolated up-spins
to flip up that is related to the size of the domains of down-spins
separating these up-spins (this will be quantified in
section~\ref{ACIC}). As the system gets closer to equilibrium these
domains become longer and the energy barriers increase. Thus the
dynamic slows down.

Dynamical constraints can also induce 
 `entropy barriers'. In this case  there are no  energy
barriers to relaxation, rather one can imagine special
configurations or small doors in the phase space that the
system must pass through to allow it to relax.  These doors are found
through random exploration of the phase space.  However, as the
relaxation proceeds the doors become progressively fewer and
harder to find.  One example of such a system is the
Backgammon Model\cite{Ritort}.

Finally one should contrast the idea of energy (or free energy)
barriers induced by dynamical constraints with energy barriers induced by
disorder. It is well known that in a disordered system, where there is
competition between different quenched random interactions, one can
have large energy barriers in the phase space. It has been argued that
such quenched disorder can mimic glassy systems (which are generally
non-disordered). The theoretical machinery developed in the study of
disordered systems then allows one to to proceed in calculations
\cite{KT}. However, one still has to come up with arguments that
relate the quenched disorder to some dynamically `self-induced
disorder' \cite{BM,MPR}. 
Alternatively one can go one step further and simply make
assumptions about the phase space, such as a valley structure, without
specifying how such features arise as a result of the microscopic
model.  This forms the basis of the `trap model'
\cite{Trap,Vilgis}.  In that scenario one has a distribution of
trap depths in the phase space.  With increasing time the system will
explore deeper and deeper traps and remain in them for longer and
longer.

\subsection{Coarsening}
\label{Intro:coarse}
We now turn to the idea of coarsening and make comparisons with the
glassy dynamics discussed so far.  To visualise a coarsening system 
think of quenching a system from a high temperature phase where its
order parameter is zero to a low temperature where the
order parameter can take some number of non-zero values
(each different value corresponding to a different ordered phase). Domains of
the ordered phase(s) emerge and grow in time and it is this phenomenon
that is referred
to as coarsening.  At late times the system
enters a {\em scaling  regime}, that is a
regime characterised by a single length scale
(the typical size of domains) $\ell(t)$ that grows as $\ell(t) \sim
t^n$. In this  late time scaling regime
the distribution of domains, once scaled by $\ell(t)$,
is statistically invariant.
Thus the typical domain size indicates the age of the coarsening
system.  The value of the exponent $n$ depends on the symmetry of the
order parameter and conservation laws of the system.  For a review see
\cite{Bray}.

As a concrete example consider the zero temperature Ising model with
Glauber spin-flip dynamics (non-conserved order parameter).  Domains
of up-spins and down spins are separated by domain walls that perform
random walks: a step of the walk corresponds to the event that one of
the spins adjacent to the domain wall flips.  When two domain walls meet
they annihilate and a domain is eliminated.  Straight away one can
argue roughly that, since the domain wall motion
is diffusive, the time typically required to eliminate domains of
size $l$ is $T(l) \sim l^2$. Thus the typical domain size after
time $t$ is $\ell \sim t^{1/2}$.

More precisely one can write  the growth law as a differential equation
by noting that the rate of change of the typical domain size should be
proportional to the inverse of the mean time to eliminate a domain
multiplied by the size of the domain being eliminated. 
\begin{equation}
\frac{\partial \ell}{\partial t}
\propto \frac{\ell}{T(\ell)} = \frac{1}{\ell}
\label{domains}
\end{equation}
and one recovers $\ell \sim t^{1/2}$. This growth
exponent  actually  holds
for a non-conserved scalar order parameter in two dimensions
and above \cite{Bray}.

Now consider generalising (\ref{domains}) to the case where some
energy barrier $\Delta E$ (or more strictly free energy
barrier) is involved in the elimination of the
domains and the system is at low but finite temperature. 
\begin{equation}
\frac{\partial \ell}{\partial t}
\propto \frac{e^{-\Delta E/T}}{\ell}\;.
\label{domain}
\end{equation}
Actually this precludes one dimensional systems that only order up to
a finite length scale at finite temperature (see Section~\ref{ABC}), but
at very low $T$ one can consider the ordering process up to that
finite length.  Some possible scenarios resulting from (\ref{domain})
have been categorised in \cite{SHS}. If the barriers 
$\Delta E$ are independent
of $\ell$ one recovers $\ell \sim t^{1/2}$ growth. If the barriers are
proportional to $\ell^m$ one obtains $\ell \sim \left[ \ln
  t\right]^{1/m}$, in particular $m=1$ yields logarithmic growth.  In
Section~\ref{ACIC} it will be shown that for the ACIC discussed in
Section~\ref{KinCon} the energy barriers are  logarithmic in $\ell$
thus yielding from (\ref{domain}) a growth law where the growth
exponent is proportional to the temperature as $T\to 0$. In
section~\ref{ABC} models will be discussed
that have energy barriers that are linear in $\ell$ thus yielding
domain growth that is logarithmic in time.
For a  particular model, the ABC model \cite{EKKM}
it will be shown how these linear energy barriers arise.
We  refer to such cases where something
different from power law growth with temperature independent exponent
is exhibited as {\em anomalous} coarsening.

A common approach in the study of coarsening is to make an
approximation of a mean-field nature.  That is, one focusses on
the probability distribution of the domain sizes and ignores spatial
correlations between domains. Such an approach
is variously referred to as an
Interparticle Distribution Function \cite{DbA}, Independent Intervals
Approximation \cite{KBN} or a `bag model' \cite{DGY}. To visualise this
one thinks of placing all the domains in a bag {\it i.e.}  forgetting how
domains are arranged with respect to one another.  Then the dynamics
become the updating of the domains in the bag. At each update a
domain is selected from the bag along with temporary neighbours.  An
update rule is implemented that depends on the shape/size of the
domain and its neighbours. After the update the domains are all
replaced in the bag.  For example, in models where domain walls
diffuse and annihilate or coalesce, such as the kinetic Ising model
discussed above, the update rule is to lengthen and shorten domains
according to how the domain wall moves. In another class of models
\cite{DGY,BDG},  domains are eliminated  at each update with
probabilities  depending on their size; then  the length of an
eliminated domain is distributed amongst the neighbours.

\subsection{Nonequilibrium steady states}
So far, although we have discussed glassy systems out of thermal
equilibrium, the dynamics have implicitly been assumed to obey detailed
balance in the equilibrium state.  Detailed balance means that in the
steady state there is no net flow of probability between any two
configurations.  However one can consider a much larger class of
nonequilibrium systems that are defined solely by their dynamics,
without reference to any energy function.  Although these systems may
relax to some steady state it need not be a steady state described by
Gibbs-Boltzmann statistical mechanics.  In general there will be a net
flow of probability between pairs of configuration, leading to
probability current loops in the configuration space.

Examples of nonequilibrium steady states are given by driven systems
with open boundaries where a mass current is driven through the
system. Thus the system is driven by its environment rather being in
thermal equilibrium with its environment.  In such a driven steady
state generic long range correlations may be exhibited \cite{SZ}. This
is in contrast to an equilibrium state which only exhibits long-range
correlations at non-generic points {\it i.e.}  phase transitions.  Since
there is no energy function and Gibbs-Boltzmann statistical mechanics
does not apply, there is no general formulation within which to solve
for such a driven steady state. However it has often been suggested
that the generic long-range correlations may result from some
effective long-range Hamiltonian that could describe the dynamics.

Of particular interest have been one-dimensional nonequilibrium
systems. For models respecting detailed balance it is well known that
no phase transition or ordering process that continues indefinitely
can occur. However for nonequilibrium systems this is not the
case\cite{David}.  Thus nonequilibrium systems afford new
possibilities for coarsening processes even in one-dimensional
systems\cite{Evans}. Moreover there are a number of exactly solvable one
dimensional nonequilibrium systems\cite{Privman}.  In
section~\ref{ABC} the steady state of the ABC model will be solved for
some special cases and it will be shown how strong phase separation
along with anomalous coarsening can occur in one dimensional systems.

\section{Asymmetric Constrained Ising Chain}
\label{ACIC}
In this section I discuss the model introduced by J{\"{a}}ckle and
Eisinger\cite{JE,EJ}. As discussed in Section~\ref{KinCon} it was
originally introduced as a model of cooperative, glassy relaxation.
In particular the directed nature of the dynamical constraint implies
a hierarchy in the spin relaxation.  As I shall now describe the
directed nature of the constraint also makes the model amenable to
analysis.

\subsection{Model definition}
The model comprises $L$ Ising  spins  $s_i=0,1$ on a one-dimensional lattice 
with periodic boundary conditions (site $i=L+1$ is
identified with site $i=1$). The dynamics are defined by the following
spin-flip rates 
\begin{equation}
\begin{array}{ccl}
1\ 1 \to 1\ 0 & \mbox{with rate} & 1\\[1ex]
1\ 0 \to 1\ 1 & \mbox{with rate} & \ds \epsilon
\end{array}\label{dynamics}
\end{equation}
where
\begin{equation}
\epsilon =\exp(-1/T)\;. \label{epsdef}
\end{equation}
Thus a spin can only flip if its left neighbour is pointing up (note
that in \cite{JE} the mirror image of the above definition was used,
so that that right neighbour had to point up for a spin to be able to
flip).  By a rate, say $x$, we mean that in a small time $dt$ the
event happens with probability $x dt$. It is easy to check that the
dynamics obey detailed balance with respect to an energy function $E =
\sum_{i=1}^L s_i$ {\it i.e.} the equilibrium distribution corresponds
to free spins in a downwards pointing field:
\begin{equation}
P_{\rm eq}(\{s_i\}) = \frac{1}{Z} \exp \left[ - \frac{\sum_i s_i}{T} \right]
= \frac{\epsilon^{\sum_i s_i}}{(1+\epsilon)^L}\;.
\label{Peq}
\end{equation}
It follows that the equilibrium concentration of up-spins, $c = \langle s_i \rangle$, is
given by
\begin{equation}
c = \frac{\epsilon}{1+\epsilon}\;.
\label{c}
\end{equation}
Although (\ref{Peq}) would hold for a number of different dynamics
obeying detailed balance, other properties such as two time
correlation functions may be sensitive to the particular choice of
dynamics.

It is clear that the asymmetric 
dynamical constraint implies that
information propagates to the right only.  Thus in a thermodynamic
limit
where information cannot propagate all the way around the ring back to
the starting point, we must have
$\langle s_{i+k}(0) s_{i}(t) \rangle_{\rm eq} -c^2 =0$
for $k>0$. However, the fact that detailed balance holds
implies that in the steady state we must have reversibility.
To see this note that when one has detailed balance there is no net
flow of probability between any two configurations. Since there
is no flow of probability there is nothing to
distinguish the forwards direction in time from backwards
direction.
Therefore running
the systems backwards in time will not change 
any two time correlation functions and
\begin{equation}
\langle s_{i+k}(0) s_{i}(t) \rangle_{\rm eq} 
= \langle s_{i+k}(t) s_{i}(0) \rangle_{\rm eq}\;.
\label{reverse}
\end{equation}
Since we have argued that for $k>0$
the left hand side of (\ref{reverse}) is equal to $c^2$
we deduce that the connected correlation function must
be site diagonal:
\begin{equation}
 \langle s_{i}(0) s_{j}(t) \rangle_{\rm eq} -c^2 
\propto \delta_{ij}\;.
\end{equation}
This result is particular to the fully asymmetrically constrained model.

\begin{figure}
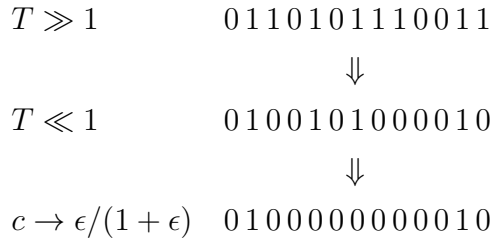

\begin{eqnarray*}
T \gg 1\quad&& 0\,1\,1\,0\,1\,0\,1\,1\,1\,0\,0\,1\,1\\
&&\hspace{1.5cm}\Downarrow\\
T \ll 1\quad&& 0\,1\,0\,0\,1\,0\,1\,0\,0\,0\,0\,1\,0\\
&&\hspace{1.5cm}\Downarrow\\
c \to \epsilon/(1+\epsilon)\quad&& 0\,1\,0\,0\,0\,0\,0\,0\,0\,0\,0\,1\,0
\end{eqnarray*}
\vspace*{0.2cm}
\caption{Schematic representation of the evolution
  of the system following a deep quench.  Before the quench $T\gg1$
  and $c$, the concentration of up-spins, $\simeq1/2$.  After the
  quench all the `mobile' up-spins (i.e. those adjacent to other
  up-spins) are eliminated first. A slow coarsening process ensues that
  reduces the density of up-spins $c$ to its equilibrium value.
 }
\label{fig:quench}
\end{figure}

We will be interested mainly in the behaviour after a quench from
equilibrium at some high initial temperature $T \gg 1$ to low
temperature $T\ll 1$ ($\epsilon \to 0$).  At low temperatures the
equilibrium concentration of up-spins (\ref{c}) is small.  Thus
the quench is followed
by a process of elimination of up-spins. However to eliminate an
up-spin one first has to generate an adjacent up-spin. This implies
energy barriers in the system's evolution.  In figure~\ref{fig:quench}
 the sequence of events after such a quench is illustrated
schematically.

The basic objects that we use for the description of the system are
{\em domains}. As shown by the vertical lines in
% the sample configuration
\[
\ldots1|0001|1|1|01|001|1|1|01|0\ldots,
\]
a domain consists of an up-spin and all the down-spins that separate it
from the nearest up-spin to the left. The length $d$ of a domain then
gives the distance between the up-spin at its right edge and the
nearest up-spin to the left. Note that adjacent up-spins are counted
as separate domains of length $d=1$. In equilibrium, the distribution
of domain lengths and its average are
\begin{equation}
P_{\rm eq}(d)=\epsilon/(1+\epsilon)^d \qquad d_{\rm eq}=1+1/\epsilon\; .
\label{pd_equil}
\end{equation}
Now consider what happens after a deep quench to $T\ll 1$, $\epsilon\ll
1$. The equilibrium concentration of up-spins at the final temperature
$T$ is $c=1/d_{\rm eq}=\epsilon+\order(\epsilon^2)$; hence the equilibrium
probability of finding an up-spin within a chain segment of {\em
finite} length $d$ is $\order(d\epsilon)$ and tends to zero for $\epsilon\to
0$. In this limit ($\epsilon\to 0$ at fixed $d$), the flipping down of
up-spins therefore becomes {\em irreversible} to leading order. In
terms of domains, this means that the coarsening dynamics of the
system is one of coalescence of domains: an up-spin that flips down
merges two neighbouring domains 
into one large domain. 

Such coarsening processes have been studied in a variety of contexts.
In particular irreversible coarsening processes in which the rate of
elimination depends solely on the domain size have a very convenient
property: during such a process, no correlations between the lengths
of neighbouring domains can build up if there are none in the initial
state~\cite{BDG}.  For the present model the equilibrated initial
state consists of domains independently distributed according to
(\ref{pd_equil}) and is indeed uncorrelated. We take advantage of this
property in Section~\ref{ACIC:coarse} to obtain an exact solution of
the coarsening dynamics. We first discuss in detail how energy
barriers arise in the dynamics.

\subsection{Energy barriers}
First we estimate the typical rate $\Gamma(d)$ at which domains of
length $d$ disappear by coalescing with their right neighbours.
Because domain coalescence corresponds to the flipping down of
up-spins, $\Gamma(d)$ can also be defined as follows. Consider an open
spin chain of length $d$, with a `clamped' up-spin ($s_0=1$) added on
the left.  Starting from the state $(s_0,s_1, \ldots, s_d)$ =
$10\ldots 01$, $\Gamma^{-1}(d)$ is the typical time needed to reach
the empty state $10\ldots 00$ where spin $s_d$ has `relaxed' {\it
  i.e.} has flipped down.  Any instance of this relaxation process can
be thought of as a path connecting the initial and final
 states. Let us call the
maximum number of `excited' spins (up-spins except $s_0$) encountered
along a path its height $h$. One might think that the relaxation of
spin $s_d$ needs to proceed via the state 11\ldots 1, giving a path of
height $d$. In fact, the minimal path height $h(d)$ is much lower and
given by
\begin{equation}
\height(d)=n+1 \quad \mbox{for}\quad 2^{n-1} < d \leq 2^n
\label{hierarchy}
\end{equation}
where $n=0, 1, \ldots$.

To get a feeling for the result (\ref{hierarchy}) consider in
figure~\ref{fig:barrier} some small domain sizes.
\begin{figure}
\[
\begin{array}{llc}
d=1\quad
&
1 \,\underline{1}  \to \,1\,0 &\quad h(1) = 1\\[1ex]
d=2\quad
&
1 \,0\, 1  \to \,1\underline{\,1 \,1} \to \,1\,1\,0 \to \,1\,0\,0  &\quad h(2) =2\\[1ex]
d=3\quad
&
1 \,0\,0\, 1  \to \,1\,1\,0 \,1 \to \,1\,\underline{1\,1\,1}\to \cdots \to \,1\,0\,0\,0  &\quad h(3) =3\\[1ex]
d=4\quad
&
1 \,0\,0\,0\, 1 \to  \cdots \to 1\,\underline{1\,1}\,0 \,\underline{1} \to
1\,0\,1\,0 \,1 \to &\\[1ex]
&\hspace{1.0cm} \,1\,0\,\underline{1\,1\,1} \to
\cdots \to \,1\,0\,1\,0\,0 \to \cdots \to 1
\,0\,0\,0\,0 &\quad h(4) =3
\end{array}
\]
\vspace*{0.2cm}
\caption{Paths through spin configurations in the 
 elimination of a  domain of size $d$  that traverse the minimum energy
 barrier. The height of the barrier is
 $h(d)$
and the highest energy configuration(s) along the path has its
excess excited spins underlined}
\label{fig:barrier}
\end{figure}
The figure illustrates that to generate an up-spin adjacent to the
right boundary spin of the domain one can proceed via a sequence of
stepping-stones {\it e.g.} for $d=4$ one first generates an isolated up-spin
in the middle of the domain then uses this stepping-stone to generate
the subsequent excited spins in a similar manner to the relaxation of
a $d=2$ domain.

The result (\ref{hierarchy}) is easily demonstrated for
$d=2^n$~\cite{MJ}. To relax the $2^n$-th spin $s_{2^n}$, one can first
flip up $s_{2^{n-1}}$ and use it as a stepping-stone for relaxing
$s_{2^n}$. The corresponding path is (with $s_{2^{n-1}}$ and $s_{2^n}$
underlined)
\begin{equation}
1\ldots\underline{0}\ldots\underline{1}   \to 
 1\ldots\underline{1}\ldots\underline{1}   \to 
 1\ldots\underline{1}\ldots\underline{0}   \to 
 1\ldots\underline{0}\ldots\underline{0} 
\end{equation}
and reaches height $h(2^n)=h(2^{n-1})+1$; the $+1$ arises because the
stepping-stone stays up while the spin $2^{n-1}$ to its right is
relaxed. Continuing recursively, one arrives at $\height(2^n) =
\height(1)+n$; but $\height(1)=1$ because the only path for the
relaxation of $s_1$ is $11\to 10$. Thus we obtain equation~\eq{hierarchy}
for $d=2^n$; a proof for general integer $d$ is given in \cite{SE}.

From \eq{hierarchy} it is evident that the energy barrier $\Delta E$
for the elimination of a domain of size $d$ is $\Delta E \simeq \ln
d/\ln 2$. Thus the rate at which such domains are eliminated is
\[\Gamma(d) \sim
\epsilon^{(-\ln d/\ln 2)} =d^{- 1/T\ln 2}\]
From the discussion of 
Section~\ref{Intro:coarse} and equation~\ref{domains}
we deduce that the typical domain size grows 
and the typical energy (number of up-spins) decreases as 
\begin{equation}
d_{\rm typ} \sim t^{T\ln2}\qquad  E_{\rm typ} \sim t^{-T\ln2}\;.
\end{equation}
Also since $d_{\rm eq} \simeq \epsilon = {\rm e}^{1/T}$ the equilibration time
is
\begin{equation}
 t_{\rm eq} \sim \exp \left[ 1/ T^2\ln2 \right]\;.
\label{EITS}
\end{equation}

\subsection{Hierarchical coarsening}
\label{ACIC:coarse}
From the scaling of $\rate(d)$, the coarsening dynamics in the limit
$\eps\to 0$ naturally divides into stages distinguished by
$n=\height(d)-1=0, 1, \ldots$. During stage $n$, the domains with
lengths $2^{n-1}<d\leq 2^{n}$ disappear; we call these the `active'
domains. This process takes place on a timescale of
$\order(\rate^{-1}(d))=\order(\eps^{-n})$; because the timescales for
different stages differ by factors of $1/\eps$, we can treat them
separately in the limit $\eps\to 0$. Thus during stage $n$ active
domains are eliminated and the distribution of inactive domains
($d>2^n$) changes because elimination of an active domain implies
coalescence with a neighbouring domain and results in the creation of
a new inactive domain.

As discussed above for the irreversible system there are no
correlations between neighbouring domains.  Therefore we can work
directly with the probability of domain sizes $P(d,t)$ i.e. the
Independent Intervals Approximation sketched in Sec~\ref{Intro:coarse}
is actually exact.

First let us write down the master equation  during stage $n$
where we assume that all domains with $d\leq 2^{n-1}$ have been eliminated.
\begin{equation}
\frac{\partial}{\partial t} \p(d,t) =
- \rate(d)\p(d,t) + \sum_{d' = 2^{n-1}+1}^{\infty }
\rate(d') \p(d',t) \p(d-d',t)\;.
\label{eqn_motion}
\end{equation} 
The first term in the right hand side of (\ref{eqn_motion})
represents domains of size $d$ being eliminated; the second term
represents the domains of size $d$ being create through a domain of
size $d'$ coalescing with a domain of size $d-d'$.
Now we introduce rescaled time
$\tau=t\eps^n$; during stage $n$ of the dynamics and in
the limit $\eps\!\to\! 0$, it can take on any positive value $\tau>0$.
Defining 
$\tilde{\rate} (d)=\rate (d)/\epsilon^n$ and taking the limit
$\epsilon \to 0$  the master equation reduces to
\begin{eqnarray}
\fl\mbox{for}\quad 2^{n} \geq d > 2^{n-1}\quad&&\ddtau \p(d,\tau) =
- \tilde{\rate}(d)\p(d,\tau) \label{actrate}\\
\fl\mbox{for}\quad d > 2^n \quad&&\ddtau \p(d,\tau) =
\sum_{d' = 2^{n-1}{+}1}^{2^{n}}
\tilde{\rate}(d') \p(d',t) \p(d-d',t)\;. \label{inactrate}
\end{eqnarray}
To  proceed we define the generating function
\begin{equation}
G(z,\tau)=\sum_{d=2^{n-1}{+}1}^{\infty} \p(d,\tau)z^d
\end{equation}
and its analogue for the active domains,
\begin{equation}
  H(z,\tau)=\sum_{d = 2^{n-1}+1}^{2^n} \p(d,\tau)z^d\;.
\end{equation}
Then multiplying (\ref{actrate},\ref{inactrate}) by $z^d$ and summing
appropriately yields
\begin{eqnarray}
\fl  \qquad\ddtau H(z,\tau) &=& - \sum_{2^{n-1}{+}1}^{2^{n}}
  \tilde{\rate}(d)\p(d,\tau) z^d \\
\fl  \qquad\ddtau G(z,\tau) &=& \qquad\ddtau H(z,\tau) +
  \sum_{d=2^{n}{+}1}^{\infty}\, \sum_{d'=2^{n{-}1}{+}1}^{2^{n}}
  \tilde{\rate}(d') \p(d',\tau) \p(d-d',\tau) z^d \nonumber\\
  &=& \qquad \left(\ddtau H(d,\tau)\right)\left[1-G(z,\tau)\right]
\label{Geqn}
\end{eqnarray}
(where the last equality follows by reordering the sums over $d$ and
$d'$).  Equation~\ref{Geqn} may be integrated and one obtains
\begin{equation}
\frac{1-G(z,\tau)}{1-G(z,0)}
=\exp-\left[H(z,\tau)-H(z,0)\right]\;.
\end{equation}
Now at the end of stage $n$, all domains that were active during that
stage have disappeared, and so $H(z,\infty)=0$. Thus
\begin{equation}
G(z,\infty)-1 = [G(z,0)-1]\exp[H(z,0)]\;.
\end{equation}

Recall that we are considering stage $n$ of the dynamics.  The initial
condition for stage $n+1$ of the dynamics will be given by the
distribution $\p(d,t)$ at the end of stage $n$.  Thus defining $G_n
\equiv G(z,0)$ for stage $n$, with a similar definition for the active
generating function $H_n$, we can relate the different stages of the
dynamics through
\begin{equation}
G_{n+1}(z)-1 = [G_n(z)-1]\exp[H_n(z)]\;.
\label{main}
\end{equation}
\begin{figure}
%
%\hspace*{-1mm}
\begin{center}
\epsfig{file=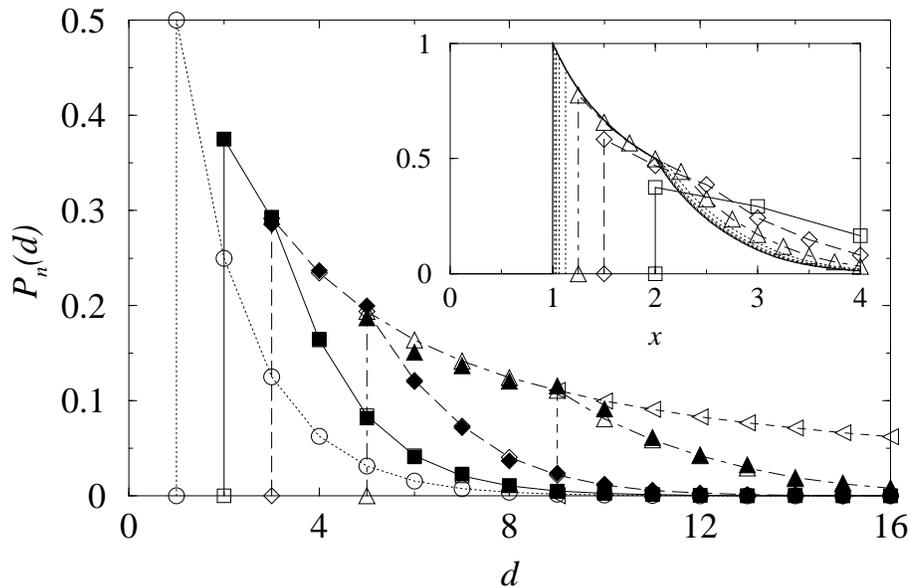, width=12cm}
\end{center}
\caption{Domain length distributions $\p_n(d)$ at the end of stage
  $n-1$ of the low $T$ coarsening dynamics, for initial temperature
  $T=\infty$. Open symbols and lines: Theoretical results, calculated
  from~(\protect\ref{main}), for $n=0$ ($\bigcirc$; initial
  condition), 1 (Boxes), 2 (diamonds), 3 (triangles). Full
  symbols: Simulation results for a chain of length $L=2^{15}$ and
  $\eps=10^{-4}$ ($n=1, 2$) and $\eps=10^{-3}$ ($n=3$). Inset: Scaled
  predictions $2^{n-1}\p_n(d=2^{n-1}x)$ vs.\ $x$ for $n=1, \ldots, 8$.
  Bold line: Predicted scaling function~(\protect\ref{scaling_limit}).
  (Figure taken from \cite{SE}).
\label{fig:pd}
}
\end{figure}
This exact result relates, through their generating functions,
$\p_n(d)$ and $\p_{n+1}(d)$ which are defined as the domain length
distributions at the end of stages $n-1$ and $n$ of the dynamics
respectively.  Iterating it from a given initial distribution
$\p_0(d)$ gives $\p_n(d)$ for all $n=1, 2, \ldots$.  Figure~\ref{fig:pd}
shows numerical results for the case where $\p_0(d)$ is the
equilibrium distribution\eq{pd_equil} corresponding to an initial
temperature of $T=\infty$.  It is clear that a scaling limit emerges
for large $n$.  By this it is meant that rescaled distributions
\begin{equation}
\ptilde_n(x)=2^{n-1}\p_n(d)
\quad\mbox{where}\quad
x=\frac{d}{2^{n-1}}\;,
\end{equation}
converge to a limiting distribution $\ptilde(x)$ for the scaled domain
size $x$.  This is just a statement of the invariance of the
coarsening processes in each stage once the domain sizes are rescaled
by the characteristic size domain size $2^{n-1}$.

The change to a continuous variable $x$ for the domain lengths simply
results in generating functions $G(z,\tau), H(z,\tau)$ being replaced
by Laplace transforms.  Invariance under\eq{main} then gives the
equation
\begin{equation}
g(2s)-1 = [g(s)-1]\exp[h(s)]\;
\label{mainLap}
\end{equation}
where
\begin{equation}
  g(s,\tau) = \int_1^\infty dx\, \ptilde(x)\, {\rm e}^{-sx}
  \qquad h(s,\tau) = \int_1^2 dx\,\ptilde(x)\, {\rm e}^{-sx} \;.
\end{equation}

We found a solution to this equation by noting that the numerics
strongly suggest $\; \ptilde(x)=1/x$ for $1<x<2$.  Using this as an
ansatz implies
\[h(x) = \mbox{Ei}(s) - \mbox{Ei}(2s)\quad\mbox{where}
\quad \mbox{Ei}(s) = \int_s^\infty\frac{{\rm e}^{-u}}{u}\,du\]
which when inserted into (\ref{mainLap}) yields
\begin{equation}
\left[1- g(s)\right] \exp (\mbox{Ei}(s)) = constant
\end{equation}
The requirement that $g(s) \to 0$ for large $s$ fixes the constant as
unity which yields
\begin{equation}
g(s) = 1-\exp(- \mbox{Ei}(s))\;.
\end{equation}
Expanding the exponential as a series allows the Laplace transform to
be inverted term by term and one obtains 
\begin{eqnarray} 
\ptilde(x)&=&
\frac{1}{2\pi i}\int_{\gamma-i\infty}^{\gamma+i \infty}
{\rm e}^{sx} g(s) \nonumber \\
&=& \frac{1}{2\pi i}\int_{\gamma-i\infty}^{\gamma+i \infty} {\rm
  e}^{sx} \sum_{m=1}^\infty (-)^{m+1} \frac{{\rm Ei}^m}{m!}
\nonumber \\
&=& \sum_{m=1}^\infty \frac{(-1)^{m+1}}{m!} \int_1^\infty\!
\prod_{r=1}^{m}\frac{dx_r}{x_r} \ 
\delta\left(\sum_{s=1}^{m}x_s-x\right) \nonumber
\\
&=&\Theta(x-1)\,\frac{1}{x}-\Theta(x-2)\,\frac{\ln(x-1)}{x}+\ldots
\label{scaling_limit}
\end{eqnarray}
where $\Theta(x)$ is the Heaviside step function. 
This series (\ref{scaling_limit}) has singularities in the $k$-th
derivative at the integer values $x=k+1$, $k+2$, \ldots

It is interesting to note that $\ptilde(x)$ given by
(\ref{scaling_limit}) is identical to the scaling function obtained
for a simple `paste-all' model of coarsening wherein the smallest
domain on a one dimensional lattice is eliminated by pasting it onto
one of its neighbours\cite{DGY}.

The calculated $\ptilde(x)$ agrees well with the results obtained by
direct iteration of\eq{main} (figure~\ref{fig:pd}).  The average domain
length in the scaling limit is given by $\dbar_n=2^{n-1}\xbar$; from
the results for $\ptilde(x)$ we find $\xbar=\exp(\gamma)=1.78\ldots$,
where $\gamma$ is Euler's constant.

\subsection{Stretched exponential relaxation}
The result (\ref{EITS}) for the EITS equilibration time
$\teq=\exp(1/T^2\ln 2)$ is based on the extrapolation of the
finite-$\dbar$ coarsening behaviour, $\dbar\sim t^{T\ln 2}$, into the
equilibrium region $\dbar= d_{\rm eq}= \order(1/\eps)$, where it is no
longer strictly valid. We now show, however, that the same timescale
is obtained from the initial decay of the spin-spin autocorrelation
function defined by
\begin{equation}
R(t) = \frac{ \langle s_i(0) s_i(t) \rangle_{\rm eq}
}{\langle s_i(0) \rangle_{\rm eq}}
\label{Rdef}
\end{equation}
{\em at equilibrium}  at
low temperature $T$. 

In equation~\ref{Rdef} $R(t)$ is the probability that an up-spin at $t=0$
is also up at a later time $t$. As $t$ increases, $R(t)$ decays from 1
to the equilibrium concentration of up-spins, $c=\eps/(1+\eps)$. To
find the initial decay of $R(t)$, consider the early stages of the
dynamics ({\it i.e.} $t=\order(\eps^{-\logtime})$ with $\logtime$
finite).  For $\logtime \to n+$, all domains of length $d\leq 2^n$
will have disappeared because $t\gg \rate^{-1}(d)$.  Therefore only
up-spins that bounded longer domains at $t=0$ will have an $\order(1)$
probability of still being up.  From the equilibrium
distribution\eq{pd_equil}, one sees that they constitute a fraction
$(1+\eps)^{-2^n}$ of the up-spins at $t=0$, hence $R(\logtime=n+0)
\simeq 1-2^n\eps+\order(\eps^2)$.  Neglecting corrections of
$\order(\eps^2)$, the quantity $-\ln R(\logtime)$ thus lies between
$2^{\logtime-1}\eps$ and $2^{\logtime}\eps$ (for $\logtime>0$).

Reverting to unscaled time $t$, we have 
\begin{equation}
{\textstyle 1/2\leq -[\ln
  R(t)]/(t/\teq)^{T\ln2} \leq 1 } 
\end{equation}
for short times $(t/\teq)^{T\ln
  2}\ll 1$ which implies
\begin{equation}
R \simeq \exp -\left[ a \left(t/t_{\rm eq}\right)^{T\ln2}\right]
\end{equation}
{\it i.e.} stretched exponential relaxation with a stretching exponent
that depends linearly on $T$.  However this argument only holds
strictly for $\left(t/t_{\rm eq}\right)^{T\ln2} \ll 1$.  For longer
times the stretched exponential behaviour no longer holds and one
requires a more sophisticated analysis \cite{SE01}.

Note that the timescale $\teq$ that enters here is
the same as the equilibration
time $\teq=\exp(1/T^2\ln 2)$ found above. Thus {\em to leading
order} we can identify the
equilibration time for coarsening after a quench, with the equilibrium
relaxation time; both have an EITS-divergence at low $T$.
However the corrections to the EITS-divergence for the two timescales
( {\it e.g.} factors of the form $\exp(a/T)$) need not be equal.

\subsection{Other related models}
Recently it has been shown that the coarsening theory described above
for ACIC is also relevant to a three-spin interaction Ising Model on a
triangular lattice \cite{NM}. There a dual model entails the
elimination of defects (corresponding to up-spins in the ACIC),
subject to dynamical constraints. An EITS relaxation time is obtained
and the Independent Intervals Approximation described here (exact for
ACIC) can be used as a good approximation to the coarsening dynamics
(see Garrahan this volume).

One can also interpolate between the asymmetric constrained model and
the symmetric constrained model\cite{FA} by introducing a parameter
$b$ into the dynamics \cite{BG}. The spin-flip rates are
\begin{equation}
\begin{array}{ccc}
1\ 1& { {\ds 1{-}b \atop \ds \longrightarrow}\atop {\ds \longleftarrow
    \atop 
\ds (1{-}b)\epsilon}} & 1\ 0 \\[1ex]
1\ 1& { {\ds b \atop \ds \longrightarrow}\atop {\ds \longleftarrow
    \atop 
\ds b\epsilon}} & 0\ 1
\end{array}
\end{equation}
The ACIC is recovered when $b=0$ and the symmetric model
is recovered when $b=1/2$. 
In the symmetric model an isolated up-spin can effectively diffuse
by creating a neighbouring up-spin then flipping down the original
up-spin. The domain coalescence
happens through this diffusion process.
Thus the energy barrier for elimination of domains
does not depend on domain size.  This implies an Arrhenius relaxation
law and `strong' glass behaviour. The crossover to the `fragile' glass
behaviour seen for the  ACIC ({\it i.e.} EITS relaxation time)
as one varies the parameter $b$ has been studied
(see Buhot this volume).

Finally let us mention a  constrained Ising spin chain where the
field is induced dynamically \cite{MDG}.  The allowed spin flips have
rates defined as follows
\begin{eqnarray}
0\,1\,1 &{ {\ds 1/2 \atop \ds \longrightarrow}\atop {\ds \longleftarrow
    \atop 
\ds 1/2}}&0\,0\,1 \nonumber \\[1ex]
1\,1\,0 &{ {\ds 1/2 \atop \ds \longrightarrow}\atop {\ds \longleftarrow
    \atop 
\ds 1/2}}& 1\,0\,0  \\[1ex]
0\,1\,0 &{ \stackrel{\ds 1}{\ds \longrightarrow}}&0\,0\,0 \nonumber
\end{eqnarray}
Thus a down-spin inside an up-spin domain cannot flip up but
an up-spin inside a down-spin domain can flip down.
The domains of up-spins grow as normal as $t^{1/2}$ but the down-spins
domains grow slightly more quickly as $t^{1/2}\ln t$. Eventually
this results in
a slow decay of the magnetisation (number of up-spins as)
$c \sim 1/\ln t$.
\section{The ABC Model}
\label{ABC}
\subsection{Coarsening in one-dimensional systems}
First let us review why indefinite coarsening does not occur in
equilibrium systems in one dimension.  Perhaps the best known argument
is that of Landau and Lifshitz \cite{LL}. (Other arguments are
summarised in \cite{Evans}.) For simplicity, consider a
one-dimensional lattice of $L$ sites with two possible states, say $A$
and $B$, for each site variable.  Let us assume the ordered phases,
where all sites take state $A$ or all sites take state $B$, have the
lowest energy, and assume a domain wall (a bond on the lattice which
divides a region of $A$ phase from that of $B$) costs a finite amount
of energy $\epsilon$.  Then $n$ domain walls will cost energy $n
\epsilon$ but the entropic contribution to the free energy due to the
number or ways of placing $n$ walls on $L$ sites $\simeq nT \left[ \ln
(n/L) -1 \right]$ $\mbox{for}\quad 1 \ll n \ll L$. Thus for any finite
temperature a balance between energy and entropy ensures that the
number of domain walls grows until it scales as $L$, that is, until
the typical ordered domain size is finite.

Note that this argument relies on a finite energy cost for domain
walls, and short-range interactions so that one may ignore the
interaction energy of domain walls.  If the domain walls can feel each
other through some long-range mechanism then coarsening can ensue
\cite{OEC}.  Also, of course, we require non-zero temperature so that
entropy comes into play.  In contrast, at zero temperature the
one-dimensional kinetic Ising model discussed in the introduction does
coarsen.

In the following we will discuss a one-dimensional model where
although the dynamics are local the systems coarsens. In a special case
one can understand this through the existence of an effective
long-range energy function.

\subsection{Model definition}
\label{Sec:ABCdef}
%----------------------------------
%%%%%%%%%%%%%%%%%%%%%%%%%%%%%%%%%%%%%%%%%%%%%%%%%%%%%%%%%%%%%%%%%%%%%%%%%%%%%
%%%%%%%%%%%%%%%%%%%%%%%%%%%%%%%%%%%%%%%%%%%%%%%%%%%%%%%%%%%%%%%%%%%%%%%%%%%%%
%%%%%%%%%%%%%%%%%%%%%%%%%%%%%%%%%%%%%%%%%%%%%%%%%%%%%%%%%%%%%%%%%%%%%%%%%%%%%
Here we define a model, to be referred to as the ABC model, that
exhibits phase separation in one dimension. Consider a one-dimensional
periodic lattice of length $N$ where each site is occupied by one of
the three types of particles, $A$, $B$, or $C$. The model evolves
under a random sequential update procedure which is defined as
follows: at each time step a pair of neighbouring sites is chosen
randomly and the particles at these sites are exchanged according to
the following rates
\begin{eqnarray}
A\,B~~{ {\ds q \atop \ds \longrightarrow}\atop {\ds \longleftarrow
    \atop 
\ds 1}}~~B\,A \nonumber \\[1ex]
B\,C~~{ {\ds q \atop \ds \longrightarrow}\atop {\ds \longleftarrow
    \atop 
\ds 1}}~~C\,B  \label{eq:dynamics} \\[1ex]
C\,A~~{ {\ds q \atop \ds \longrightarrow}\atop {\ds \longleftarrow
    \atop 
\ds 1}}~~A\,C \nonumber \;.
\end{eqnarray}
\noindent The particles thus diffuse asymmetrically around the ring. The
dynamics conserve the number of particles, $N_A , N_B$ and $N_C$ of
the three species.

The $q=1$ case is special. Here the diffusion is symmetric and every
local exchange of particles takes place with the same rate as the
reverse move. The system trivially obeys detailed balance reaching a
steady state in which all microscopic configurations (compatible with
the number of particles $N_A , N_B$ and $N_C$) are equally probable.
This state is disordered and homogeneous; no phase separation takes
place.

\begin{figure}[h]
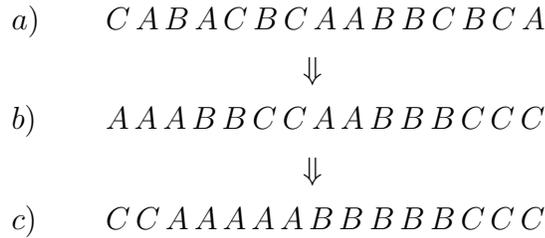

\begin{eqnarray*}
a)\qquad&& C\,A\,B\,A\,C\,B\,C\,A\,A\,B\,B\,C\,B\,C\,A\\
&&\hspace{2.5cm}\Downarrow\\
b)\qquad&& A\,A\,A\,B\,B\,C\,C\,A\,A\,B\,B\,B\,C\,C\,C\\
&&\hspace{2.5cm}\Downarrow\\
c)\qquad&&C\,C\,A\,A\,A\,A\,A\,B\,B\,B\,B\,B\,C\,C\,C
\end{eqnarray*}
\vspace*{0.1cm}
\caption{Schematic representation of the evolution
  of the system starting from a random initial condition a).
  Initially all unstable domain walls ($B\,A$, $C\,B$ or $A\,C$) are
  removed and one arrives at a metastable state b).  A slow coarsening
  process ensues in which the smallest domains are eliminated
  until one arrives at a fully phase separated state c).  Compare with
  figure~\ref{fig:quench}.  }
\label{fig:quenchABC}
\end{figure}
Now consider the case $q<1$ (the case $q > 1$ can easily be understood
by symmetry).  As a result of the bias in the exchange rates an $A$
particle moves preferentially to the left inside a $B$ domain and to the
right inside a $C$ domain. Similarly the motion of $B$ and $C$
particles in foreign domains is biased. Consider the dynamics starting
from a random initial configuration figure~\ref{fig:quenchABC} a). The
configuration is composed of a random sequence of domains of $A$, $B$,
and $C$ particles. Due to the bias a local configuration in which an
$A$ domain is placed to the right of a $B$ domain is unstable and the
two domains exchange places on a relatively short time scale which is
linear in the domain size.  Similarly, $AC$ and $CB$ domains are
unstable too. On the other hand $AB$, $BC$ and $CA$ configurations are
stable and long-lived. Thus after a relatively short time the system
reaches a state of the type illustrated in figure~\ref{fig:quenchABC} b)
in which $A,B$ and $C$ domains are located to the right of $C,A$ and
$B$ domains, respectively. The evolution of this state takes place via
a slow diffusion process in which, for example $A$ particles have to
diffuse against a drift over an adjacent $B$ domain.  The time scale
for an $A$ particle to cross is $\propto q^{-l}$, where $l$ is the
size of the $B$ domain. We use the discussion of
Section~\ref{Intro:coarse} and equation~\ref{domains} to deduce that the
system coarsens with an average domain size that increases with time
as $\ln t/\vert \ln q\vert$.  Eventually the system phase separates
into three domains of the three species of the form $A\ldots AB\ldots
BC\ldots C$.

In a finite system the phase-separated state may further evolve and
become disordered due to fluctuations.  However, the time scale for
this to happen grows exponentially with the system size. For example
it would take a time of order of $q^{-{\rm min}\{N_B,N_C\}}$ for the
$A$ domain in the totally phase separated state to break up into
smaller domains. Hence in the thermodynamic limit, this time scale
diverges and the phase separated state remains stable provided the
density of each species is non-zero. Note that there are always small
fluctuations about a totally phase separated state. However, these
fluctuations affect the densities only near the domain boundaries.
They result in a finite width for the domain walls (the density
profile is not a step function but is smeared out like a Fermi
function). The fact that any phase-separated state is stable for a
time exponentially long in the system size amounts to a breaking of
the translational symmetry {\it i.e.} there are $N$ equivalent ground
states but the system has to spontaneously choose one of them.

Since the exchange rates are asymmetric, the system generically
supports a particle current in the steady state which implies that
detailed balance does not hold.  To see this, consider the $A$ domain
in the phase separated state.  An $A$ particle near the $\ldots AB
\ldots$ boundary can traverse the entire $B$ domain to the right with
an effective rate proportional to $q^{N_B}$.  Once it crosses the $B$
domain it will move through the $C$ domain with speed $1{-}q$. Similarly
an $A$ particle near the $\ldots CA \ldots$ boundary can traverse the
entire $C$ domain to the left with an effective rate proportional to $q^{N_C}$.
Once the domain is crossed it moves through the $B$ domain with speed
$1{-}q$. Hence the net $A$ particle current is of the order of $q^{N_B}
- q^{N_C}$. Since this current is exponentially small in system size,
it vanishes in the thermodynamic limit.  For the case of
$N_A=N_B=N_C$, this argument suggests that the current is strictly
zero for any $N$.

The arguments presented above suggesting phase separation for $q<1$
may be easily extended to $q>1$. In this case, however, the phase
separated state is $BAC$ rather than $ABC$. This may be seen by noting
that the dynamical rules are invariant under the transformation $q
\rightarrow 1/q$ together with $A \leftrightarrow B$.

%%%%%%%%%%%%%%%%%%%%%%%%%%%%%%%%%%%%%%%%%%%%%%%%%%%%%%%%%%%%%%%%%%%%%%%%%%%%%
\subsection{Special Case $N_A=N_B=N_C$}
%%%%%%%%%%%%%%%%%%%%%%%%%%%%%%%%%%%%%%%%%%%%%%%%%%%%%%%%%%%%%%%%%%%%%%%%%%%%%
%%%%%%%%%%%%%%%%%%%%%%%%%%%%%%%%%%%%%%%%%%%%%%%%%%%%%%%%%%%%%%%%%%%%%%%%%%%%%
%%%%%%%%%%%%%%%%%%%%%%%%%%%%%%%%%%%%%%%%%%%%%%%%%%%%%%%%%%%%%%%%%%%%%%%%%%%%%
The general argument presented in the previous subsection suggests
that for the special case $N_A=N_B=N_C$, in the steady state ({\it
  i.e.} after the coarsening process) there are no currents for any
system size. We demonstrate this explicitly by showing that the {\it
  local dynamics} of the model satisfies detailed balance with respect
to a {\it long-range asymmetric} energy function ${\cal H}$.

We define the occupation variables as follows:
\begin{eqnarray}
\label{ocup}
A_i = \left\{ \begin{array}{ll}
            1 & \mbox{if site $i$ is occupied by an $A$ particle} \\
            0 & \mbox{otherwise. }
            \end{array}
       \right.
\end{eqnarray}
The variables $B_i$ and $C_i$ are defined similarly. Clearly the
relation $A_i+B_i+C_i = 1$ is satisfied.  It turns out that for the
case $N_A=N_B=N_C=N/3$ the steady-state distribution $W_N(\{X_i\})$
corresponding to the dynamics (\ref{eq:dynamics}) may be written in
terms of an energy function ${\cal H}$:
\begin{equation}
\label{eq:weight}
W_N(\{X_i\}) = Z_N^{-1}q^{{\cal H}(\lbrace X_i \rbrace)} \;.
\end{equation}
\begin{equation}
\label{eq:Hamilton}
{\cal H}(\lbrace X_i \rbrace) =
\sum_{i=1}^{N-1}\sum_{j=i+1}^{N}[C_iB_j+A_iC_j+B_iA_j]-(N/3)^2 \; .
\end{equation}
Here $Z_N$ is the partition sum given by $\sum q^{{\cal H} (\lbrace
  X_i \rbrace)}$, where the sum is over all configurations in which
$N_A=N_B=N_C$.  Note that although the system is periodic and site 1
is not in any way special (\ref{eq:Hamilton}) appears to single out
site 1. Thus it is not clear that (\ref{eq:Hamilton}) is
translationally invariant under relabelling of the spins.

In order to turn equation (\ref{eq:weight}) into a usual Boltzmann form one
could define $q$ as a temperature variable with
\begin{equation}
kT=-1/\ln q\;.
\label{qtemp}
\end{equation}
Thus, $q \to 1$ is the infinite-temperature limit, corresponding to
the disorder state where each configuration is equally likely.  The
proof of equations~(\ref{eq:weight},\ref{eq:Hamilton}) is straightforward.
This is done by considering a nearest-neighbour particle exchange and
verifying that detailed balance is satisfied with respect to
(\ref{eq:dynamics}). Then we just have to check that the energy
function is translationally invariant.  We defer the proof to
Section~\ref{ABCgen} where we consider a more general $m$-species
model.

Before proceeding further to evaluate the partition sum associated
with the energy function (\ref{eq:Hamilton}) let us make a few
observations. The ground state of the energy function is given by the
fully separated state $A\ldots AB\ldots BC\ldots C$ and its
translationally related states.  It easy to check that the ground
state with the $A$ domain beginning at site 1 has zero energy since
the contribution to the sum in (\ref{eq:Hamilton}) (coming from the
$A_i\, C_j$ term) is equal to $(N/3)^2$.  A simple way of evaluating
the energy of an arbitrary configuration is obtained by noting that
nearest-neighbour exchanges $AB \rightarrow BA, BC \rightarrow CB$ and
$CA \rightarrow AC$ cost one unit of energy each while the reverse
exchanges result in an energy gain of one unit. The energy of an
arbitrary configuration may thus be evaluated by starting with the
ground state and performing nearest neighbour exchanges until the
configuration is reached, keeping track of the energy changes at each
step of the way.  The highest energy is $N^2/9$ and it corresponds to
the totally phase separated configuration $A\ldots AC\ldots CB\ldots
B$ and its $N$ translations.  Note that the majority of configurations
have energy proportional to $N^2$. In Section~\ref{Sec:mss} it will be
shown that this implies that only the ground states and low energy
excitations about them contribute to the equilibrium state.

To write ${\cal H}$ in a manifestly translationally invariant form we
define ${\cal H}_{i_0}(\lbrace X_i \rbrace)$ as the energy function in
which site $i_0$ is the origin. Namely,
\begin{equation}
\label{hi0}
{\cal H}_{i_0}(\lbrace X_i \rbrace) = \sum_{i=i_0}^{N+i_0-2}
\sum_{j=i+1}^{N+i_0-1}[C_iB_j+A_iC_j+B_iA_j] -(N/3)^2 \; ,
\end{equation}
where the summation over $i$ and $j$ is modulo $N$. Summing
(\ref{hi0}) over all $i_0$ and dividing by $N$, one obtains,
\begin{equation}
{\cal H}(\{ X_i \})   = 
 \sum_{i=1}^N \sum_{k=1}^{N-1}
(1 - \frac{k}{N}) (C_i B_{i+k} + A_i C_{i+k} + B_i A_{i+k}) - (N/3)^2 \; ,
\label{eq:H_tinvar}
\end{equation}
where in the summation the value of the site index $(i+k)$ is modulo
$N$. In the energy function (\ref{eq:H_tinvar}) the interaction is
linear in the distance between the particles, and thus is long ranged.
The distance is measured in a preferred direction from site $i$ to
site $i+k$.  Thus the interaction is asymmetric.

%%%%%%%%%%%%%%%%%%%%%%%%%%%%%%%%%%%%%%%%%%%%%%%%%%%%%%%%%%%%%%%%%%%%%%%%%%%%
\subsection{Ground States and Metastable States}
%%%%%%%%%%%%%%%%%%%%%%%%%%%%%%%%%%%%%%%%%%%%%%%%%%%%%%%%%%%%%%%%%%%%%%%%%%%%%
\label{Sec:mss}
A minimum of the energy (\ref{eq:Hamilton}) is realised by a
configuration with no unstable domain walls ($B\,A$, $C\,B$ or $A\,C$)
so that any exchange of nearest neighbour particles results in an
increase in the energy.  As well as the $N$ ground states there are
many metastable states.  Any metastable state is composed of a
sequence of domains separated by $AB,BC$ and $CA$ domain walls {\it
i.e.} $A,B$ and $C$ domains follow $C,A$ and $B$ domains, respectively
(see figure~\ref{fig:quenchABC} b).  Therefore each metastable state
has an equal number of domains of each type.  We shall refer to any
metastable state with $s$ domains of each type, with $s=1,...,N/3$, as
an $s$-state; the total number of domains in an $s$-state is $3s$. The
$s=1$ case corresponds to the ground state while $s=N/3$ corresponds
to the $ABCABC\ldots ABC$ state, composed of a total of $N$ domains
each of length $1$.  Note that in general the domains of an $s$-state
need not be of equal length.

In the coarsening process it is these metastable states that control
the dynamics. We discuss here some properties of the states such as
their number and energy.

To obtain a bound for ${\cal N} (s)$, the number of metstable states
with $s$ domains of each species, note that the number of ways of
dividing $N/3$ $A$ particles into $s$ domains is ${{N/3 -1} \choose
  {s-1}}$.  The number of ways of combining $s$ divisions of each of
the three types of particles is clearly $\left[{N/3-1}\choose
  {s-1}\right]^3$. There are at most $N$ ways of placing this string
of domains on a lattice to obtain a metastable state (the number of
ways need not be equal to $N$ since the string may possess some
translational symmetry). One therefore has
\begin{equation}
\left[{N/3-1}\choose {s-1}\right]^3 \le{\cal N}(s) \le N
\left[{N/3-1}\choose {s-1}\right]^3\; .
\label{nMSS}
\end{equation}
Thus, the total number of metastable states is exponential in $N$. 

We now consider the energy of the metastable states. It is easy to
convince oneself that among all $s$-states, none has energy lower than
the configuration illustrated in figure~\ref{fig:mss}.
\begin{figure}[h]
\[
A\ldots AB\ldots BC\ldots CABCABC\ldots ABC 
\]
\vspace*{0.1cm}
\caption{
  Metastable state that has the lowest energy for a given number $s$
  of domains of each species.  The $3(s-1)$ rightmost domains are of
  size $1$ and the three leftmost domains are of size $(N/3-s+1)$
  each.  }
\label{fig:mss}
\end{figure}

The energy of this state, $E_s$ satisfies the following recursion
relation
\begin{equation}
\label{RR}
E_{s+1} = E_{s} + N/3 - s 
\end{equation}
with $E_1 =0$. To see this from figure~\ref{fig:mss}, note that the
$s+1$-state may be created from the $s$-state by first moving a $B$
particle from the leftmost $B$ domain across $(N/3-s)$ $C$ particles
to the right, costing $(N/3-s)$ units of energy.  Then move an $A$
particle from the leftmost $A$ domain to the right across the adjacent
$B$ and $C$ domains; no net energy change results.  Thus the total
energy cost of the moves is $(N/3-s)$, yielding (\ref{RR}). The
recursion relation (\ref{RR}), together with $E_1=0$, is readily
solved to give
\begin{equation}
E_s = (s-1) \frac{N}{3} - \frac{s(s-1)}{2} \; .
\label{eMSS}
\end{equation}
The energy of all metastable $s$-states is larger or equal to $E_s$ as
given by equation (\ref{eMSS}). Note that $E_s$ increases with $s$.
Furthermore, for finite $s$ the energy is linear in $N$ whereas for
$s\propto N$ the energy becomes quadratic in $N$.

Now let us consider the contribution of the metastable states to the
partition sum. Multiplying the upper bound on the number of $s$-state
(\ref{nMSS}) with the lower bound on the energy (\ref{eMSS}) one
obtains an upper bound on the contribution to the partition sum that
vanishes in the thermodynamic limit
\begin{equation}
\fl  N \left[{N/3-1}\choose {s-1}\right]^3 
q^{ (s-1) N/3 - s(s-1)/2 }
\to 0\quad \mbox{for $s>1$, $q<1$ and $N \to \infty$}\;.
\label{partmss}
\end{equation}
However, the contribution from the ground states $s=1$ is $N$.  Thus
even though metastable states dominate the dynamics, they do not
contribute to the partition sum since in (\ref{partmss}) the energy
grows more strongly than the entropic contribution.

%%%%%%%%%%%%%%%%%%%%%%%%%%%%%%%%%%%%%%%%%%%%%%%%%%%%%%%%%%%%%%%%%%%%%%%%%%%%%
\subsection{Partition Sum}
%%%%%%%%%%%%%%%%%%%%%%%%%%%%%%%%%%%%%%%%%%%%%%%%%%%%%%%%%%%%%%%%%%%%%%%%%%%%%
We now analyse in more detail the behaviour of the partition sum.  In
principle one wants to compute
\begin{equation}
Z_N =\sum q^{{\cal H} (\lbrace
X_i \rbrace)}
\end{equation}
where the sum is over {\em all} configurations in which $N_A=N_B=N_C$.
First note that any configuration that contains unstable domain walls
({\it i.e.} is not a metastable state) can be associated with a
metastable state by a path of decreasing energy comprising nearest
neighbour exchange eliminating of the unstable walls. Conversely the
sum over all configurations may be implemented by summing over all
ground states and metastable states and the excitations about those
states. It is not hard to believe that as the metastable states make
vanishing contributions to $Z_N$ so do excitations about them.  This
is proven rigorously in \cite{EKKM}. In the following we just consider the
excitations about the ground states.
\begin{figure}
\begin{equation}
\underline{A}\,A\cdots\,A\,\underline{A\,B\,A\,B\,A\,B}\,B\cdots \,B\,
\underline{B\,C}\,C\cdots\,C\,\underline{C}\
\end{equation}
\caption{
  Excitation about a ground state at the $A\,B$ domain wall of energy
  $E=3$: one $A$ particle has penetrated two exchanges into the $B$
  domain and a second $A$ particle one exchange. This corresponds to
  the partition 3 = 2+1. The domain wall regions are underlined.}
\label{fig:excited}
\end{figure}
Consider figure~\ref{fig:excited} where a low energy excitation about a
ground state is illustrated. The excitation is localised near the
$A\,B$ domain wall and comprises one or more $A$ particles penetrating
into the $B$ domain (equivalently $B$ particles penetrating into the
$A$ domain).  The energy cost is given by the sum of the distances
each $A$ particle has penetrated into the $B$ domain.  Thus the total
number of excitations of energy $m$ at the boundary is the number of
ways of dividing $m$ into an ordered set of integers corresponding to
the distance the first $A$ has moved, the distance the second $A$ has
moved and so on.  This is equal to $P(m)$ the number of {\em
  partitions} of the integer $m$.

Several results concerning
partitions are known\cite{Andrews}.
First the generating function of $P(m)$ is given by
\begin{eqnarray}
\sum_{m=0}^{\infty} q^m P(m)&=& \frac{1}{(q)_{\infty}}
\label{partgen}\\[1ex]
\mbox{where}\qquad
(q)_\infty &=& \lim_{n\to \infty}
(1-q)(1-q^2)\ldots (1-q^n)\;.
\label{qinfdef}
\end{eqnarray}
Although a simple explicit formula for $P(m)$ does not exist the
asymptotic behaviour is given by
\begin{equation}
P(m) \simeq \frac{1}{4m \sqrt{3}} \exp{(\pi (2/3)^{1/2} \ m^{1/2})}\;.
\end{equation}
Note that the increase is a stretched exponential in $m$
{\it i.e.} slower than exponential.

In the thermodynamic limit one can use (\ref{partgen}) directly to
calculate the sum of excitations around a domain wall {\it i.e.} the
sum over $m$ of $q^m$, the weight of an excitation of energy $m$,
multiplied by $P(m)$, the number of such excitations.  For a finite
system there should be some upper limit on $m$---for example, an $A$
particle moved across the $B$ domain will eventually reach the $C$
domain---but this upper limit can be safely taken to infinity for
large $N$ \cite{EKKM}.  By the same token the three domain walls have
no significant interaction.  Then one has that in the large $N$ limit
and for all $q<1$, the partition sum is given by
\begin{equation}
Z_N = N/[(q)_{\infty}]^3 \; .
\label{psum}
\end{equation} 
Here the factor $N$ is a result of the sum of contributions
from the $N$ ground  states and the cubic power comes from
the product of excitations at the three domain walls.

Note that the partition sum is linear and not
exponential in $N$,
meaning that the free energy is not
extensive. This reflects the fact that excitations
are localised near the domain walls.
In turns this
stems
from the fact that the energies of most configurations
are $\order(N^2)$ which is a result of the long-range interaction in the
energy function.

A consequence of this is that in the steady state the system is fully
phase separated, that is,
each of the domains is pure. This was
demonstrated
in \cite{EKKM} by showing that
\begin{equation}
\langle A_i A_{i+r} \rangle = \frac{1}{3}- \order(r/N).
\end{equation}
for any given $r$ and sufficiently large $N$.
Thus the probability of
finding a particle a large distance inside a domain of particles of
another type is vanishingly small in the thermodynamic limit.

For $q$ close to $1$, $(q)_{\infty}$ as defined in
(\ref{qinfdef}), has an essential singularity
\begin{equation}
(q)_{\infty}=\exp\left\{-\frac{1}{\ln q} \left[ \pi^2/6 + {\cal O}(1-q)
 \right]\right\}\;.
\end{equation}
This suggests that extensivity of the free energy could be restored in
the double limit $q \rightarrow 1$ and $N \rightarrow \infty$ with $N
\ln q$ finite. 
Physically one can understand this scaling variable as the ratio
of the domain length ($N/3$) to the  domain wall width
($\sim \int l q^l dl / \int q^l dl = 1/\vert \ln q \vert$).
The validity of $N \ln q$ as a scaling variable was investigated
in \cite{EKKM} where a good
scaling collapse was obtained for the
two point correlation functions.

\subsection{Coarsening}
The analytic results of the previous subsection for $N_A=N_B=
N_C$ give proof of the coarsening into three pure domains
in that special case. Clearly one expects the same behaviour
in the general case but to demonstrate it numerically
requires prohibitively long time scales.
In order to study   anomalously slow
coarsening dynamics numerically
one can employ an effective or toy model that may be more easily simulated.
In this subsection I will outline how this can be implemented.

We consider a system at time $t$ such that the average domain size,
$\langle l \rangle$, is much larger than the domain wall width. At
these time scales, the domain walls can be taken as sharp and we may
consider only events which modify the size of domains. This means that
the dynamics of the system can be approximated by considering only the
movement of particles between neighbouring domains of the same
species. Thus only metastable states are considered
in the toy model.

We represent a configuration by a sequence of integers of the form
$a_1 b_1 c_1 a_2 b_2 c_2 \ldots a_3 b_3 c_3$ where, for example, the
$i$th domain of $A$ particles is of length $a_i$.  At each time step a
pair of neighbouring domains of the same species of particle, say
$a_i$ and $a_{i+1}$, is chosen randomly.
The
exchange of particles between domains, 
takes place at a rate dictated by the size of the 
domains $b_i$ and $c_i$ which separate 
them.
Thus the lengths of
the chosen domains are modified by carrying out one of the
following processes:
\begin{equation}
  \label{toy_rule}
  \begin{array}[]{ll}
    \left. 
     \begin{array}[]{ll}
    1) & a_i \rightarrow a_i - 1   \\
      & a_{i+1}\rightarrow a_{i+1} + 1  \\
     \end{array} \right\} & {\rm with~rate} \quad q^{b_i} \\
& \\
    \left.
     \begin{array}[]{ll}
    2) & a_i \rightarrow a_i + 1    \\
      & a_{i+1}\rightarrow a_{i+1} - 1  \\  
  \end{array} \right\} & {\rm with~rate} \quad q^{c_i} \;.
\end{array}
\end{equation}
If $a_i$ becomes zero, one deletes the domain $a_i$ from the
list of domains, and merges $b_i$ and $c_i$ with
$b_{i-1}$ and $c_{i-1}$, respectively.

To simulate the toy model efficiently, an algorithm suitable for rare
event dynamics must be used due to the small rate of events
\cite{rare}. In \cite{EKKM} an algorithm was employed
that entails  repeating the
following steps:

\begin{enumerate}
\item List all possible events $\{n\}$ and assign to them
rates $\{r_n\}$ according to the rules of the model.
\item Choose an event $m$ with probability  $r_m/R$ where $R=\sum_n r_n$.
\item Advance time by $t \rightarrow t + \tau $, where $\tau =1/R\;$.
\end{enumerate}
The algorithm would be equivalent to a usual Monte Carlo simulation,
where one time step is equivalent to one Monte Carlo sweep, if in step
$3$, $\tau$ were to be drawn from a Poisson distribution $R ~\exp[-R
\tau ]$. However, a saving in computer time
can be had by making the approximation 
$\tau =1/R$.

\begin{figure}[t]
\center{\psfig{figure=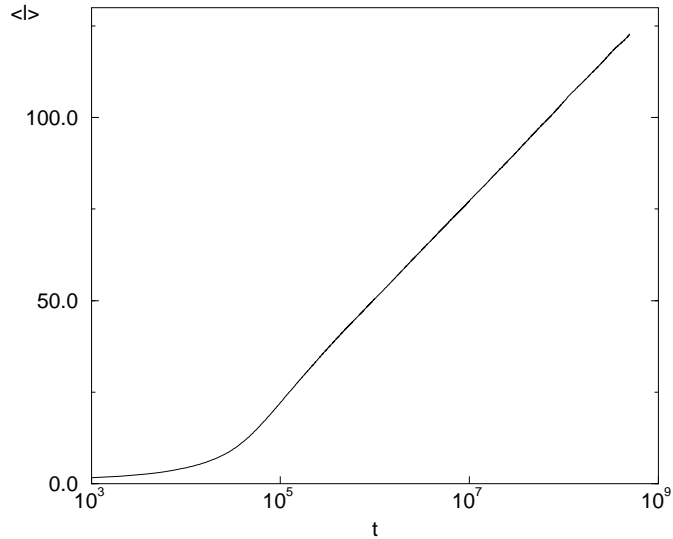,width=10cm}}
\caption{Monte Carlo simulation results for the toy model for the
 average domain size, $\langle l \rangle$, vs. time, $t$, for 
$N=9000$ and $q=0.8$. The data is averaged over 1760 runs.
There is clear evidence that $\langle l \rangle$ grows as
$\ln t$.
(Figure taken from \cite{EKKM}.) }
\label{fig_domsize}
\end{figure}

In \cite{EKKM}
 the dynamics were simulated for lattice sizes up to $9000$. For
simplicity we consider the case $N_A=N_B=N_C$. An example of  typical
behaviour of the average domain size is shown in
figure \ref{fig_domsize}. One can see that after an initial
transient growth time the data fits very well with a $\ln t$
behaviour. (Note that the
system size is large enough that the growth is $N$ independent.)
Simulations for different $q$ values indicate that,
\begin{equation}
\label{loga}
\langle l \rangle = a \ln t/ \vert \ln q \vert
\end{equation}
with $a \simeq 2.6$. The toy model enables one to verify the scaling
behaviour (\ref{loga}) and estimate the constant $a$. This would be very
difficult to do by simulation of the full model
(\ref{eq:dynamics}).

One can further analyse the toy model by using 
an Independent Intervals Approximation as discussed in
Section~\ref{Intro:coarse}. This was carried out in \cite{EKKM}.

\subsection{Generalisation to $M \geq 3$ species}
\label{ABCgen}
We now generalise the ABC model to $M$ species where $M \ge 3$.  We
may define the most general $M$ species model with  nearest neighbour
particle exchanges that conserve the number of each species as
follows.  Let $X_i=1,2,\ldots,M$ denote which type of particle is
present at site $i$: $X_i=m$ means that site $i$ is occupied by a
particle of type $m$.  Nearest-neighbour exchanges occur with the
following rates:
\begin{equation}
\label{eq:Mspeci}
mn~~{ {\ds q(m,n) \atop \ds \longrightarrow}\atop {\ds \longleftarrow
    \atop 
\ds q(n,m)}}~~nm \; ,
\end{equation}
and we take $q(m,m)=1$. The model conserves $N_m$, the number of particles
of type $m$, for all $m$.

According to the choice of the rates (\ref{eq:Mspeci}) the model may
or may not phase separate.  It is not difficult to choose rates so
that phase separation does indeed occur. For example for $M=4$ the
choice
\begin{eqnarray}
A\,B~~{ {\ds q \atop \ds \longrightarrow}\atop {\ds \longleftarrow
    \atop 
\ds 1}}~~B\,A \qquad&&
D\,A~~{ {\ds q \atop \ds \longrightarrow}\atop {\ds \longleftarrow
    \atop 
\ds 1}}~~A\,D\nonumber\\[0.5ex]
B\,C~~{ {\ds q \atop \ds \longrightarrow}\atop {\ds \longleftarrow
    \atop 
\ds 1}}~~C\,B \qquad&&
A\,C~~{ {\ds q \atop \ds \longrightarrow}\atop {\ds \longleftarrow
    \atop 
\ds 1}}~~C\,A
\label{eq:four}
\\[0.5ex]
C\,D~~{ {\ds q \atop \ds \longrightarrow}\atop {\ds \longleftarrow
    \atop 
\ds 1}}~~D\,C \qquad&&
D\,B~~{ {\ds q \atop \ds \longrightarrow}\atop {\ds \longleftarrow
    \atop 
\ds 1}}~~B\,D\nonumber
\end{eqnarray}
leads to phase separation into pure domains ordered $ABCD$.
Generally, for $M>3$ the structure of the metastable states can become quite
complicated \cite{EKKM}. For example, domains
ordered $ACDABCB$
are also metastable in the model
defined by (\ref{eq:four}).
We now find the conditions under which
the dynamics (\ref{eq:Mspeci})
satisfy detailed balance
with respect to a steady-state weight analogous to
(\ref{eq:weight},\ref{eq:Hamilton}):
\begin{equation}
  \label{eq:sweight}
  W(\{X_i\}) = \mbox{const.}\times\prod_{i=1}^{N-1} \prod_{j=i+1}^N q(X_j,X_i) \; ,
\end{equation}
where the constant is arbitrary.

Consider a particle exchange between sites $k$ and $k+1$, where
$X_k=m,X_{k+1}=n$ and $k\ne N$.
Expanding the product in (\ref{eq:sweight}), it is easy to verify that
\begin{equation}
  \frac{W(X_1,\ldots,m,n,\ldots,X_N)}{W(X_1,\ldots,n,m,\ldots,X_N)}  =
 \frac{q(n,m)}{q(m,n)} \; .
\end{equation}
Since this holds for any $m,n$, and is irrespective of the number of
particles of each species, the dynamics (\ref{eq:Mspeci})
satisfy detailed balance with respect to the weight
(\ref{eq:sweight})
for all nearest-neighbour exchanges between
 sites $k$ and $k+1$ with $k\neq N$.
If the weights (\ref{eq:sweight}) are translationally invariant
then detailed balance will also hold for exchanges between sites $1$
and $N$.

Thus, to complete the proof of detailed balance it is sufficient to
demand that (\ref{eq:sweight}) is translationally invariant. To do
this we relabel sites $i \rightarrow i+1$. The weight then becomes
\begin{equation}
 \label{eq:sweight2}
  W(\{X_i\}) = \prod_{i=1}^{N-1} \prod_{j=i+1}^N q(X_{j-1},X_{i-1}) \; ,
\end{equation}
where $X_0$ is identical to $X_N$.
Rewriting this equation by relabelling the indices we obtain,
\begin{equation}
 \label{eq:sweight3}
  W(\{X_i\}) = \left[\ \prod_{i=1}^{N-1} \prod_{j=i+1}^{N} q(X_{j},X_{i})\
 \right] \
  \prod_{k=1}^{N-1}\frac{q(X_k,X_N)}{q(X_N,X_k)} \; .
\end{equation}
Comparing (\ref{eq:sweight3}) with (\ref{eq:sweight}) and noting for
example that,
\begin{equation}
\prod_{j=1}^{N}q(X_j,X_N) = \prod_{l=1}^{M} \left[ q(l,X_N) \right]^{N_l} \; ,
\end{equation}
one can see that (\ref{eq:sweight}) is translationally invariant if
\begin{equation}
\label{condition}
 \prod_{l = 1}^M \left[\frac{q(m,l)}{q(l,m)}
                     \right]^{N_{l}} = 1 \; , 
\end{equation}
for every $m=1,\ldots,M$. Thus, detailed balance holds if
(\ref{condition}) is satisfied. 

In particular for   the 3 species model with particles labelled
$A\,B\,C$ the condition (\ref{condition}) becomes
\begin{eqnarray}
 \left[\frac{q(A,B)}{q(B,A)}
                     \right]^{N_{B}}  
 \left[\frac{q(A,C)}{q(C,A)}
                     \right]^{N_{C}} &=& 
 \left[\frac{q(B,A)}{q(A,B)}
                     \right]^{N_{A}}  
 \left[\frac{q(B,C)}{q(C,B)}
                     \right]^{N_{C}} \nonumber \\[1ex]
&=& \left[\frac{q(C,A)}{q(A,C)}
                     \right]^{N_{A}}  
 \left[\frac{q(C,B)}{q(B,C)}
                     \right]^{N_{B}}= 1\;.
\label{ABCcondition}
\end{eqnarray}
The $ABC$ model defined in Sec.~\ref{Sec:ABCdef} has $\ds
\frac{q(B,A)}{q(A,B)}=\frac{q(C,B)}{q(B,C)}=\frac{q(A,C)}{q(C,A)} = q$
in which case (\ref{ABCcondition}) reduces to $N_A=N_B=N_C$.

The ABC model has been generalised to two dimensions \cite{KBEM} where
the coarsening process generates striped domains perpendicular to the
direction of the drive. The $\ln t$ growth of these domains is
retained.

\subsection{Other related Models}
A model closely related to the ABC model is that first introduced by
Lahiri and Ramaswamy \cite{LR} in the context of sedimenting colloidal
crystals.  This model comprises two interpenetrating sublattices. On
each sublattice there reside two species of particles with no holes.
On the first sublattice (at integer sites) the particles are denoted
by $+$ and $-$ whereas on the second sub-lattice (at half-integer
sites) the particles are denoted as `tilts' $\backslash$ and $/$, for
reasons that will become apparent below.  The dynamics are defined as
\begin{eqnarray}
+\,\backslash\,-~~{ {\ds r_1 \atop \ds \longrightarrow}\atop {\ds
    \longleftarrow \atop \ds r_2}}~~-\,\backslash\,+ &\qquad& 
+\,/\,-~~{
  {\ds r_2 \atop \ds \longrightarrow}\atop {\ds \longleftarrow \atop
    \ds r_1}}~~-\,/\,+ \nonumber  \\
\label{LRdynamics}\\
/\,-\,\backslash~~{ {\ds p_2 \atop \ds \longrightarrow}\atop {\ds
    \longleftarrow \atop \ds p_1}}~~\backslash\,-\,/ &\qquad& 
/\,+\,\backslash~~{ {\ds p_1 \atop \ds \longrightarrow}\atop {\ds
    \longleftarrow \atop \ds p_2}}~~\backslash\,+\,/  \nonumber
\end{eqnarray}
{\it i.e.} the particles make nearest neighbour exchanges on their
respective
sublattices but the rates are influenced by what kind
of particle is occupying the intermediate site on the other sublattice.
In \cite{LR} more general rates than (\ref{LRdynamics}) were originally considered but
the above  dynamics appear to capture all the generic
behaviour\cite{LBR,RBDB}.

In the regime $r_1 >r_2$ and $p_1 >p_2$ phase separation is observed
into ordered domains of $\backslash -$, $\backslash +$, $/ +$, $/ -$.
That is, on each lattice the particles ultimately separate into pure
domains with one ring rotated by $\pi/2$ with respect to the other.
During the coarsening process domain lengths grow logarithmically in
time.  In \cite{LBR} the separation into pure domains, as first
analysed in the ABC model, was termed `strong phase separation'.

In the special case where each lattice is half-filled with particles
and
\begin{equation} 
\frac{r_2}{r_1}=\frac{p_2}{p_1}\equiv q
\end{equation}
(where the parameter $q$ has been introduced to allow comparison with
the ABC model), the steady state weight $W_N$ satisfies detailed
balance with respect to a long-range energy function very similar in
form to (\ref{eq:Hamilton})
\begin{equation}
W_N = Z_N^{-1}q^{{\cal H}} 
\end{equation}
where
\begin{equation}
{\cal H} =
\frac{1}{2} \sum_{i=1}^{N}\sum_{j=1}^{k} \tau_{j-1/2}\,\sigma_i \; .
\end{equation}
where $\tau$ and  $\sigma$ are each Ising spins variables for one of
the sublattices:
$\sigma = 1 (-1)$ corresponds to 
a $+$ ($-$) particle and 
$\tau = 1 (-1)$ corresponds to tilt
$/$ ($\,\backslash\,$).

A very useful intuitive picture of this energy function is to regard
$\sum_{j=1}^{k}\tau_{j-1/2}$ as the height of an interface (relative
to some origin)\cite{LBR}. Thus the dynamics correspond to the $+$ and
$-$ particles making nearest neighbour interchanges on a landscape
(implied by the tilt variables) that at the same time is evolving in a
way coupled to the $+$, $-$ particles. 
The long-range energy corresponds to the gravitational potential
energy of the $+$ particles on this landscape.
A $+$ particle will have its
energy minimised at the bottom of a  valley
of the interface.  A ground state then
corresponds to the configuration that allows the $+$ particles to
minimise collectively  their energy, namely,
the tilt variables form one deep valley
with the $+$ particles residing at the bottom:
\[
-\,\backslash\,-\,\backslash\,-\,\backslash\,-\,\backslash\,
+\,\backslash\,+\,\backslash\,+\,\backslash\,
+\,/\,+\,/\,+\,/\,+\,/\,
-\,/\,-\,/\,-\,/\,
\]
The picture of particles moving on an evolving landscape is very
appealing in the context of glassy dynamics.  It is evocative of the
glassy regime of a hard-sphere colloid where particles move in cages
formed by the other particles---when a particle escapes from its cage
it will cause the cages of other particles to be restructured.

In the regime $r_1 >r_2$ and $p_1 < p_2$ the system is in a disordered
phase. In the regime $r_1 >r_2$ and $p_1 = p_2$ the fluctuations in
the landscape are uncoupled to the particle dynamics, yet the system
still exhibits an interesting coarsening dynamics\cite{DB}.

A further model related to the ABC model
 has been studied  by Arndt, Heinzel
and Rittenberg \cite{AHR}. It was originally couched in terms of $+$,$-$
particles and holes diffusing on a one-dimensional periodic lattice
with hop rates
\begin{eqnarray}
+\,- &{ {\ds q \atop \ds \longrightarrow}\atop {\ds
    \longleftarrow \atop \ds 1}}& -\,+\nonumber \\[1ex]
+\:0 &\stackrel{\ds 1}{\ds \longrightarrow}& 0\,+\label{AHR}\\
0\:- &\stackrel{\ds 1}{\ds \longrightarrow}& -\:0 \nonumber
\end{eqnarray}
In order to make a comparison with the ABC model we identify 
a $+$ particle
with an $A$,
a $-$ particle
with a $B$,
and a hole with a $C$ then the dynamics (\ref{AHR}) become
\begin{eqnarray}
A\,B~~{ {\ds q \atop \ds \longrightarrow}\atop {\ds \longleftarrow
    \atop 
\ds 1}}~~B\,A \nonumber \\[1ex]
C\,B~~\stackrel{\ds 1}{\ds \longrightarrow}~~B\,C \\
A\,C~~\stackrel{\ds 1}{\ds \longrightarrow}~~C\,A\nonumber
\end{eqnarray}
Thus this model corresponds to the ABC model with some exchanges
forbidden and for $q<1$ one has the same strong phase separation.
However for $q>1$ this model enters a disordered phase, whereas in the
ABC model one has phase separation but with the order of domains
permuted to $ACB$. Originally it was thought that there was also a
`mixed' phase in the model \cite{AHR}.  It now appears that this is a
very strong finite size effect for $q\gtrapprox 1$
\cite{RSS}.  An appealing feature of this model is that the steady
state can, in principle, be computed
exactly by a matrix product approach
for all numbers of particles \cite{AHR}.

It should also be noted that a model with cyclic symmetry
and non-conserving dynamics that exhibits coarsening has been studied
\cite{FKB}.

\section{Conclusions}
In this paper I have reviewed a variety of simple models that exhibit
anomalous coarsening in the sense that the dynamics slow down with
time and the coarsening becomes anomalously slow---slower 
than the usual power law growth of domain size with time. In the models
discussed the slowdown in the coarsening is due to dynamical
constraints rather than any quenched disorder. The dynamical
constraints imply that energy barriers must be surmounted in order for
the system to coarsen further. Thus the system is delayed in
metastable states for increasingly long times during the coarsening
process and in this sense the evolution of the system is glassy.

In Sections~2 and 3 I focussed on two models: the ACIC and the ABC
model.  In the ACIC model the energy barriers encountered during
coarsening are logarithmic in the domain size whereas for the ABC
model the barriers are linear. This leads to the domain growth $\ell
\sim t^{T\ln 2}$ (as $T\to 0$) for the ACIC and $\ell \sim \ln t$ for
the ABC model.  Despite similarities there are distinctions between
the two models. For the ACIC the energy function is trivial---free
spins in a field---and the dynamics obey detailed balance. But it is
the existence of forbidden spin-flips that generates the energy
barriers in the relaxation paths. On the other hand in the ABC model the
dynamics are prescribed without regard to an energy function and no
exchanges (that conserve the particle numbers) are forbidden.  However
in a special case one can identify an effective long-range energy
function and the anomalous coarsening
can be explained in terms of linear energy barriers in the coarsening
process.  Lastly, the ABC model will ultimately coarsen into pure
domains whose size is related to the system size whereas the ACIC only
coarsens up to a length $1/\epsilon$ that is independent of system
size.  For both models there exist other related models that exhibit
similar behaviour, thus reinforcing their interest.

The question remains as to how faithfully the anomalous coarsening
scenario describes a true (experimental) glassy system. Although the
models are by no means meant to represent any particular system, one
recovers the correct phenomenology---for example in the ACIC we
derived stretched exponential decay of an auto-correlation function
and EITS law for equilibration.  Nevertheless it is widely thought
that a coarsening system is distinct from a glassy system.

There are several reasons for this. Firstly, in a glassy system such
as hard-sphere colloid, the ordered phase (crystalline state) is
thought to be  irrelevant since its nucleation barrier is too large. Thus there are
no coarsening domains of an ordered crystalline phase.  On the other
hand at present we do not have any known spatial order parameter for a glassy
phase so it is unclear how domains of the glassy phase would coarsen.
Secondly, coarsening implies a definite direction in the dynamics,
towards the fully coarsened state.  Contrary to this, phenomenological
trap models of glassy dynamics \cite{Trap} rely on a random
exploration of the traps that exist in the phase space. That is, when
a system manages to escape from one trap it falls randomly into
another trap. In this way, as the time scale increases, the system
will locate deeper traps and stay in them for longer.  Finally
dynamically constrained models by nature rely on specific dynamics
rather than the nature of an energy function (if it exists).  This
contrasts with the `inherent structures' approach to glassy
dynamics where the energy landscape is the key feature
\cite{Stillinger,CRRS}.  This point is examined in the present volume by
Crisanti and Ritort.

Indeed a criterion has been proposed to distinguish between coarsening
dynamics and glassy dynamics\cite{BBM} in a microscopic model.  One
runs a simulation for a certain time then makes two copies of the
simulation. These copies start from identical initial conditions
(where the initial simulation was halted) but use different
realisations of the noise in their dynamics {\it i.e.}  different sets
of random numbers in a Monte-Carlo simulation.  Then, if the states of
the two simulations remain strongly correlated (have a finite overlap)
as time goes by one has a coarsening system. This is termed type I
behaviour. On the other hand, if the states of the two simulations
become less and less correlated then the system is a glassy system
this is termed type II behaviour. The idea behind this relies on the
belief that for a coarsening system there is always a preferred
direction in the phase space along which all simulations will be
swept, whereas in a glassy system the traps in the phase space are
essentially explored randomly by each simulation.

At present it is not clear how general this categorisation is.  For
example it is not yet clear how systems that coarsen but have many
ground states, such as the ABC model or the model of \cite{LBR}, are
accounted for.  

Let us also mention a 3-d ferromagnetic Ising model
with plaquette interactions\cite{Lipowski}.  In this model one has the
usual doubly degenerate ferromagnetic ground states.  In addition,
flipping any plane of spins does not increase the energy. Thus in
total there is an exponential number of degenerate ground states. In
a quench to low temperature type II glassy behaviour is
exhibited\cite{SBTB} yet it is thought that  coarsening 
occurs whereby a characteristic length grows anomalously slowly,
possibly as $\ln t$.  For a related model of competing ferromagnet and
next nearest neighbour antiferromagnetic interactions that has just
the two ferromagnetic ground states, logarithmic coarsening is
observed and explained in terms of energy barriers\cite{SHS}.  The
glassy coarsening dynamics of the plaquette model is not so well
understood\cite{LJ}.
It would certainly be of interest to broaden our understanding 
by studying further examples of anomalous coarsening.

\ack 
It is a pleasure to thank my collaborators Yariv Kafri, Hari M.
Koduvely, David Mukamel, Peter Sollich with whom the work of Sections
2 and 3 was carried out. I would also like to thank M. Barma, R. A.
Blythe, M. E. Cates, M. Clincy, A. Crisanti, D. Das, S. N. Majumdar,
F. Ritort and A. Rocco for stimulating discussions.

\end{document}